\documentclass{article}

\PassOptionsToPackage{numbers, compress}{natbib}

\usepackage[dblblindworkshop, final]{neurips_2025}
\workshoptitle{The Second Workshop on GenAI for Health: Potential, Trust, and Policy Compliance}

\usepackage[utf8]{inputenc} 
\usepackage[T1]{fontenc}    
\usepackage{hyperref}       
\usepackage{url}            
\usepackage{booktabs}       
\usepackage{amsfonts}       
\usepackage{nicefrac}       
\usepackage{microtype}      
\usepackage{xcolor}         
\usepackage[pdftex]{graphicx}
\usepackage{subcaption} 
\usepackage{diagbox}
\title{Hearing Health in Home Healthcare: Leveraging LLMs for Illness Scoring and ALMs for Vocal Biomarker Extraction}

\author{
\textbf{Yu-Wen Chen}$^{1}$\thanks{These authors contributed equally. \texttt{\{yu-wen.chen, william.ho\}@columbia.edu}.
} \quad
\textbf{William Ho}$^{1}$\footnotemark[1] \quad
\textbf{Sasha M. Vergez}$^{3}$ \quad
\textbf{Grace Flaherty}$^{3}$ \quad \\
\textbf{Pallavi Gupta}$^{2}$ \quad
\textbf{Zhihong Zhang}$^{2}$ \quad
\textbf{Maryam Zolnoori}$^{2}$ \\ 
\textbf{Margaret V. McDonald}$^{3}$ \quad
\textbf{Maxim Topaz}$^{2}$ \quad
\textbf{Zoran Kostic}$^{1}$ \quad
\textbf{Julia Hirschberg}$^{1}$ \\
$^{1}$The Fu Foundation School of Engineering and Applied Science, Columbia University \\
$^{2}$School of Nursing, Columbia University \\
$^{3}$Center for Home Care Policy \& Research, VNS Health 
}

\begin{document}

\maketitle

\begin{abstract}
The growing demand for home healthcare calls for tools that can support care delivery. In this study, we explore automatic health assessment from voice using real-world home care visit data, leveraging the diverse patient information it contains. First, we utilize Large Language Models (LLMs) to integrate Subjective, Objective, Assessment, and Plan (SOAP) notes derived from unstructured audio transcripts and structured vital signs into a holistic illness score that reflects a patient’s overall health. This compact representation facilitates cross-visit health status comparisons and downstream analysis. Next, we design a multi-stage preprocessing pipeline to extract short speech segments from target speakers in home care recordings for acoustic analysis. We then employ an Audio Language Model (ALM) to produce plain-language descriptions of vocal biomarkers and examine their association with individuals’ health status. Our experimental results benchmark both commercial and open-source LLMs in estimating illness scores, demonstrating their alignment with actual clinical outcomes, and revealing that SOAP notes are substantially more informative than vital signs. Building on the illness scores, we provide the first evidence that ALMs can identify health-related acoustic patterns from home care recordings and present them in a human-readable form. Together, these findings highlight the potential of LLMs and ALMs to harness heterogeneous in-home visit data for better patient monitoring and care.

\end{abstract}

\section{Introduction}
The global demand for home healthcare highlights the urgent need for improved tools and interventions to enhance care delivery. Automatic health assessment systems have the potential to assist clinicians by identifying aspects that may be overlooked during visits and support individuals with limited access to clinical services. Recently, Large Language Models (LLMs) have undergone development at an unprecedented pace, demonstrating contextual understanding, reasoning, and task generalization across diverse domains~\citep{brown2020language, dong2022survey, wei2022chain}. These advancements are increasingly being leveraged in the field of healthcare, with LLMs being applied to tasks such as summarizing and extracting structured information from clinical notes, answering medical questions, and facilitating patient-provider communication~\citep{agrawal2022large, singhal2025toward, van2024adapted, wen2024leveraging, wiest2024privacy}. However, to the best of our knowledge, no prior study has explored the use of LLMs to interpret and assess a patient's overall health condition from real-world home care visits. Such data is inherently heterogeneous and complex, comprising both structured and unstructured information~\citep{zolnoori2025beyond}. For instance, Subjective, Objective, Assessment, and Plan (SOAP) notes~\citep{podder2020soap} capture diverse aspects of a patient’s condition in narrative form, while vital signs~\citep{sapra2023vital} consist of multiple numerical indicators reflecting distinct physiological states.

Recognizing the strength of LLMs in generating human-readable text and leveraging extensive textual knowledge, researchers have extended LLMs into Audio-Language Models (ALMs)~\citep{chu2024qwen2, ghosh2025audio, hurst2024gpt, tang2024salmonn}. ALMs inherit LLMs’ abilities to generate interpretable descriptions and knowledge of human voices, while also possessing the capability to extract meaningful information directly from audio signals. While ALMs have demonstrated strong performance in areas such as audio captioning and sound event classification~\citep{deshmukh2023pengi}, their applications in health assessment are largely underexplored -- despite growing evidence that acoustic signals can provide valuable indicators of health conditions~\citep{baur2024hear}, such as respiratory symptoms~\citep{xia2021covid} and depression~\citep{leal2024speech}.

In this study, we explore the performance of LLMs in assessing real-world home care visit data. We address the challenges of comparing heterogeneous data across visits by consolidating them into a single, holistic illness severity score (denoted as illness score) that reflects a patient’s overall health status for each visit. This approach is motivated by evidence that LLMs encode extensive medical knowledge and can reason effectively about clinical tasks~\citep{agrawal2022large,kung2023performance, lievin2024can, liu2023large, singhal2023large}, and have the potential as automatic evaluators, generating pseudo-labels that offer a scalable and resource-efficient alternative to human assessment~\citep{chen2025judgelrm, chiang2023closer, li2024llms, liu2023g, mizumoto2023exploring, zhu2025judgelm}. While not a replacement for detailed clinical documentation, this unified score provides a more compact and comparable representation of health status, enabling downstream analyses such as investigating associations of vocal biomarkers with patient health. 

We then apply ALMs to analyze acoustic signals from real-world clinician-patient conversations, investigating whether the differences in vocal biomarkers detected by ALMs correspond to patients’ overall health conditions. Specifically, we perform multi-stage processing to extract short speech segments of the target speaker from the full conversation recording and then prompt an ALM to describe the input speech in terms of various vocal biomarkers, such as voice quality, pronunciation, and energy level. In contrast to commonly used speech features such as speech embeddings from acoustic models, which cannot be directly interpreted by humans, the ALM’s ability to generate plain-language descriptions provides clinicians with more accessible insights into a patient’s condition.

The contributions of this study are summarized as follows:
\begin{enumerate}
\item Using data from real-world home care visits, our study captures authentic patient–clinician interactions to examine whether, and how, characteristics of a person’s voice are correlated with their overall health condition.
\item We propose using LLMs to integrate SOAP notes and vital signs into a compact numeric health score, serving as pseudo-labels that enable more effective and quantitatively driven downstream analyses. Experimental results demonstrate LLMs' capabilities to generate reliable illness scores, which show strong alignment with actual patient clinical outcomes.
\item To the best of our knowledge, we have conducted the first study employing ALMs to assess human health through speech signals. We show that ALMs capture acoustic patterns reflecting changes in health status and provide interpretable descriptions that facilitate clearer understanding for clinicians.
\end{enumerate}

\section{Analysis Framework}
\begin{figure}[t]
    \centering
  \includegraphics[width=\linewidth]{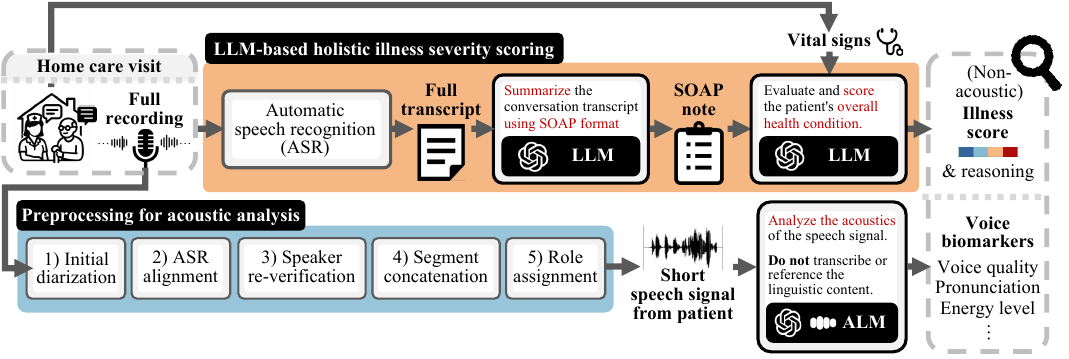} \hfill
  \caption{Overview of our analysis framework. Home care visit audio recordings are processed through two pipelines. \textbf{Top}: An LLM summarizes ASR transcripts into SOAP notes and combines them with vital signs to produce an illness score. \textbf{Bottom}: Preprocessed patient speech segments are analyzed by an ALM to generate acoustic descriptors, which are correlated with the illness score.}
    \label{fig:analysis_framework}
\end{figure}

Our goal in this study is to investigate the correlation between people’s voices and their health conditions using real-world home care visit data. First, we propose using an LLM to generate a numeric illness score to summarize the patient’s overall health status based on both SOAP notes from clinician-patient conversation transcript and recorded vital signs. The numeric illness score functions as a pseudo-label, providing a quantitative representation of each visit’s health status, making it easier to compare patient conditions over time or across individuals. Next, we extract short speech signals from a designated speaker in the conversation. We then apply an ALM to generate interpretable acoustic descriptions from the speech signals. Finally, we examine the relationship between these acoustic descriptions and the illness scores. Figure~\ref{fig:analysis_framework} outlines the analysis framework for this study.

\subsection{Real-World Home care Visit Data}
The data utilized in this study was collected from real-world home care visits. 
Participants were enrolled in a home care service, either following hospitalization or through community referral, in which clinicians conducted home visits and recorded their interactions and conversations with the patients.
All participants were instructed to maintain their usual interaction patterns and to disregard the presence of recording equipment to minimize behavior alterations. Data collected during each visit includes full audio recordings and  patient's vital signs. Audio recordings were made using wireless microphones attached to both the clinician and the patient at the start of the visit. Each vital sign -- including body mass index (BMI), blood pressure (diastolic and systolic), height, weight, pulse, respiration rate, temperature, oxygen saturation, blood sugar level, and pain level -- may be measured once, multiple times, or not at all, depending on the clinician's judgment. 

\subsection{LLM for Holistic Illness Severity Scoring}
We apply an automatic speech recognition (ASR) system to generate transcripts from the full audio recordings. The transcripts are then processed by an LLM to produce SOAP notes \citep{chen2024exploring, podder2020soap}. The SOAP notes, combined with the patient’s vital signs, are then fed to the LLM to generate an illness score along with a rationale for the score. Scoring ranges from 1 to 5, with 1 indicating good overall health and 5 indicating a critical condition for which hospitalization is recommended. We evaluate our illness scores by examining their alignment with real-world patient outcomes. Health conditions are inherently complex, making it challenging to obtain a definitive ground-truth score that represents the overall health status. One of the best available indicators of a home care patient’s overall health is whether they were hospitalized or required an emergency department (ED) visit. Accordingly, we examine whether illness scores generated by LLMs are associated with patients’ risk of being hospitalized or visiting the ED within a 60-day home care period. We categorize patients into two groups: 1) those with at least one ED visit or hospitalization during this period (denoted as ED/HOSP), and 2) those without (denoted as No ED/HOSP). For each patient, we calculate the average illness score, then compare these averages between the two groups. Experimental results (Section~\ref{sec:illness_score_results}) show that patients who were hospitalized or visited the ED during the period had notably higher illness scores on average, indicating that the score can reflect overall health conditions. Therefore, we use these scores as health indicators in our acoustic analysis, serving as pseudo-labels to group patients by health severity and enabling comparisons of vocal characteristics across different health levels.

\subsection{ALM for Vocal Biomarker Extraction}
\subsubsection{Preprocessing for Acoustic Analysis}

To analyze each speaker’s acoustic characteristics, we locate and extract their speech segments from raw multi-speaker recordings. Because our recordings were collected in real-world home care environments, variability from third-party speakers and background noise often cause current diarization models to mis-assign speaker identities. We therefore design a multi-stage preprocessing pipeline to automatically exclude segments with ambiguous speaker labels or mixed voices.

Only the first 30 seconds of each processed audio recording are used for acoustic analysis, since the length of processed signals can vary depending on the duration and quality of the original recordings, a fixed short segment reduces variability and ensures more consistent and comparable analysis across visits. Because processing long inputs with ALMs is computationally expensive, we begin by evaluating shorter speech signals to determine if they can effectively capture useful acoustic patterns. Finally, limiting input duration constrains the available linguistic context, reducing the risk of ALMs relying on semantic information in ways that violate the instructions.

The multi-stage preprocessing pipeline includes:
\begin{enumerate}
  \item \textbf{Initial diarization} assigns generic speaker labels to speech segments using a diarization model.
  \item \textbf{ASR alignment} applies an ASR model to obtain word-level timestamps and map them to diarization results. Words are grouped into sentences based on end-of-sentence punctuation, and any sentences containing words with unassigned speakers are removed. 
  \item \textbf{Speaker re-verification} reduces possible speaker confusion by using speaker embeddings from speech-enhanced audio to confirm diarization labels. For each diarized (generic) speaker, we compute their average embedding from all associated segments. The cosine similarity is computed between each segment embedding and all speaker’s average embeddings, assigning the most similar speaker to the segment. Segments whose newly inferred speaker deviates from the diarizer’s label are discarded. 
  \item \textbf{Segment concatenation} merges verified speech segments into continuous speech with short silences inserted between them.   
  \item \textbf{Role assignment} leverages recognizable interaction patterns in clinician-patient conversations~\citep{zolnoori2023patient}, e.g., clinicians asking patients condition-related questions, to classify each speaker as clinician, patient, or third party. An LLM performs the classification based on full transcripts, and manual inspection of a subset confirms high accuracy in role assignment. 
\end{enumerate}

\subsubsection{Analysis of ALM Vocal Biomarkers and Health}
For acoustic analysis, we evaluate the ALMs' ability to detect basic speaker traits, including gender and age, as baseline tasks to measure the model’s capacity to extract speaker information not represented in the transcript. We then investigate whether the ALM’s acoustic descriptions of \emph{emotion}, \emph{voice quality},  \emph{pronunciation characteristics (fluency, prosody, articulation)},  \emph{energy level}, and  \emph{potential signs of discomfort or fatigue} reflect variations in an individual’s health condition. We then analyze speech signals from clinicians vs. patients based on the hypothesis that these two groups differ in overall health status. Finally, we explore how the ALM’s acoustic descriptions of patient voices evolve with health status indicated by the illness scores.

\section{Experimental Process}
\subsection{Data Collection and Statistics}
The data used in this study were collected under Institutional Review Board (IRB) approval. Both patients and clinicians are compensated for their participation. All textual data were anonymized prior to our access. However, since audio recordings inherently reflect speaker identity information, all experiments were conducted within a secure, access-controlled computing environment to ensure data privacy and protection. Additionally, only open-source software or HIPAA-compliant commercial services (i.e., the HIPAA-compliant version of OpenAI’s API) were used. 

The dataset comprises single-channel (48 kHz) audio recordings from 160 home care visits (one recording per visit), ranging from approximately 6 to 70 minutes in duration, with an average length of about 30 minutes. At least one vital sign was recorded during each visit. The dataset includes 77 patients (23 male, 54 female, ages 20–99) and 9 clinicians, with no demographic information collected for clinicians. For each patient, we derived a binary outcome indicating whether an ED visit or hospitalization occurred within 60 days of the start of home care -- 20 patients experienced at least one such event. If a patient returned home after an ED visit or hospitalization during this period, clinicians could continue recording, and these post-event recordings were included. Although enrollment periods varied, all visit data was collected within the first 60 days after the start of care,  spanning from June 16, 2024, to May 6, 2025.

\subsection{Model settings}
Illness scores and SOAP notes were generated independently for each visit. We compare the performance of \emph{GPT-4.1}~\citep{achiam2023gpt, openai2025gpt41} with default settings against open-source LLMs: \emph{Qwen-3-8B}~\citep{yang2025qwen3} and \emph{Llama-3.1-8B-Instruct}~\citep{dubey2024llama}, which serve as open-source alternatives to GPT-4.1, as well as \emph{MediPhi-Instruct}~\citep{corbeil2025modular} chosen for its specialized medical expertise. Detailed prompts are in Appendix~\ref{sec:appendix_prompts}.

We used \emph{nvidia/parakeet-tdt-0.6b-v2}~\citep{parketeernvidia} as the ASR model for both full recording transcription and the preprocessing for acoustic analysis. 
Because data primarily focuses on conversations between the patient and clinician, we selected \emph{nvidia/diar\_sortformer\_4spk-v1}~\citep{parksortformer} as the initial diarization model due to its strong performance in few-speaker settings. However, since the full recordings are significantly longer than the training data, we split each recording into 250-second chunks with a 5-second overlap, and processed each chunk independently. Speakers were matched across chunks by extracting embeddings using the speaker verification model \emph{nvidia/speakerverification\_en\_titanet\_large}~\citep{koluguri2022titanet}, which were then clustered into four groups via agglomerative hierarchical clustering. For speaker re-verification, SEMamba~\citep{chao2024investigation} with the \emph{vd.pth} checkpoint was employed for speech enhancement, followed by verification using the same speaker verification model as above. Sentences that fail verification are discarded. Verified sentences longer than 0.5 seconds are concatenated with 0.5 seconds of silence inserted between them, and the first 30 seconds are retained for analysis. GPT-4.1 with default settings is used for role assignment in acoustic processing. The ASR, speech diarization, and speaker verification models used in this study are publicly available on Hugging Face. For ALMs, we utilized \emph{gpt-4o-audio-preview} ~\citep{hurst2024gpt} configured with default settings and \emph{Qwen2-Audio-7B-Instruct}~\cite{chu2024qwen2}.

\section{Results}
\subsection{LLM-based Holistic Illness Severity Scoring}\label{sec:illness_score_results}

\begin{figure}[htpb!]
    \centering
    \textbf{GPT-4.1}
    \smallskip
    
    \begin{subfigure}[t]{0.29\textwidth}
        \centering
        \includegraphics[width=\textwidth]{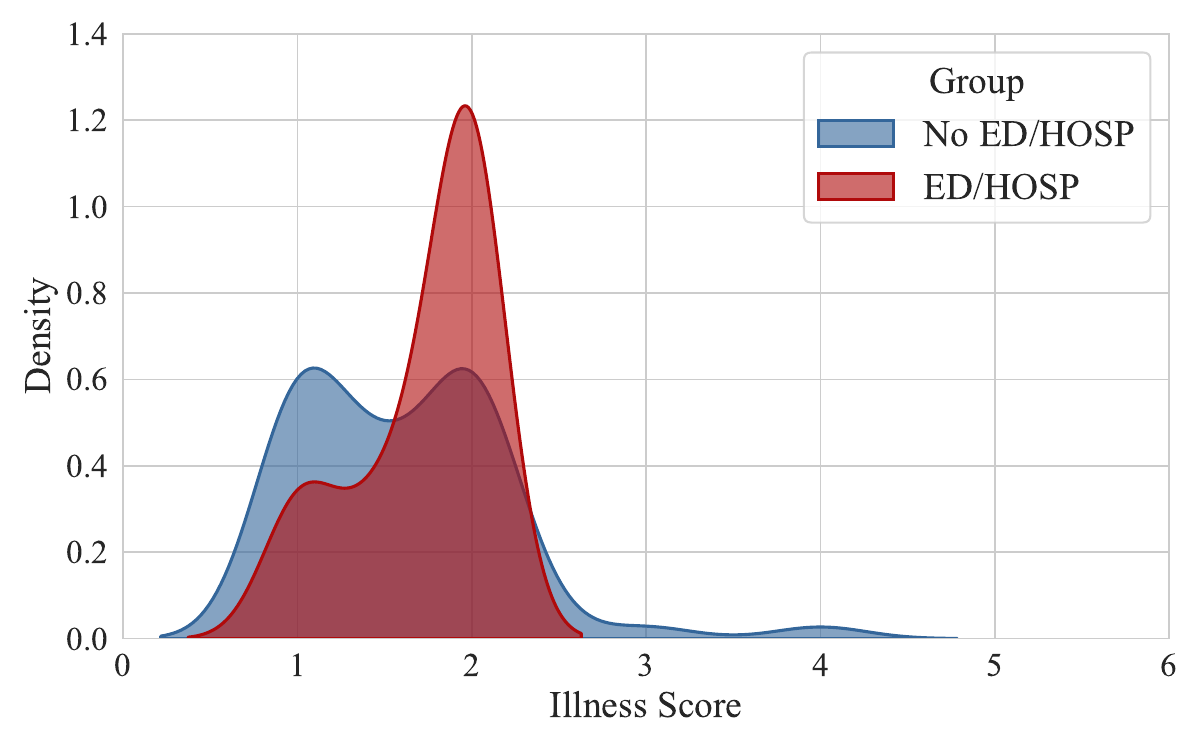}
        \caption{VITAL only}
        \label{fig:gpt41_kde_vitals_only}
    \end{subfigure}
    \hfill
    \begin{subfigure}[t]{0.29\textwidth}
        \centering
        \includegraphics[width=\textwidth]{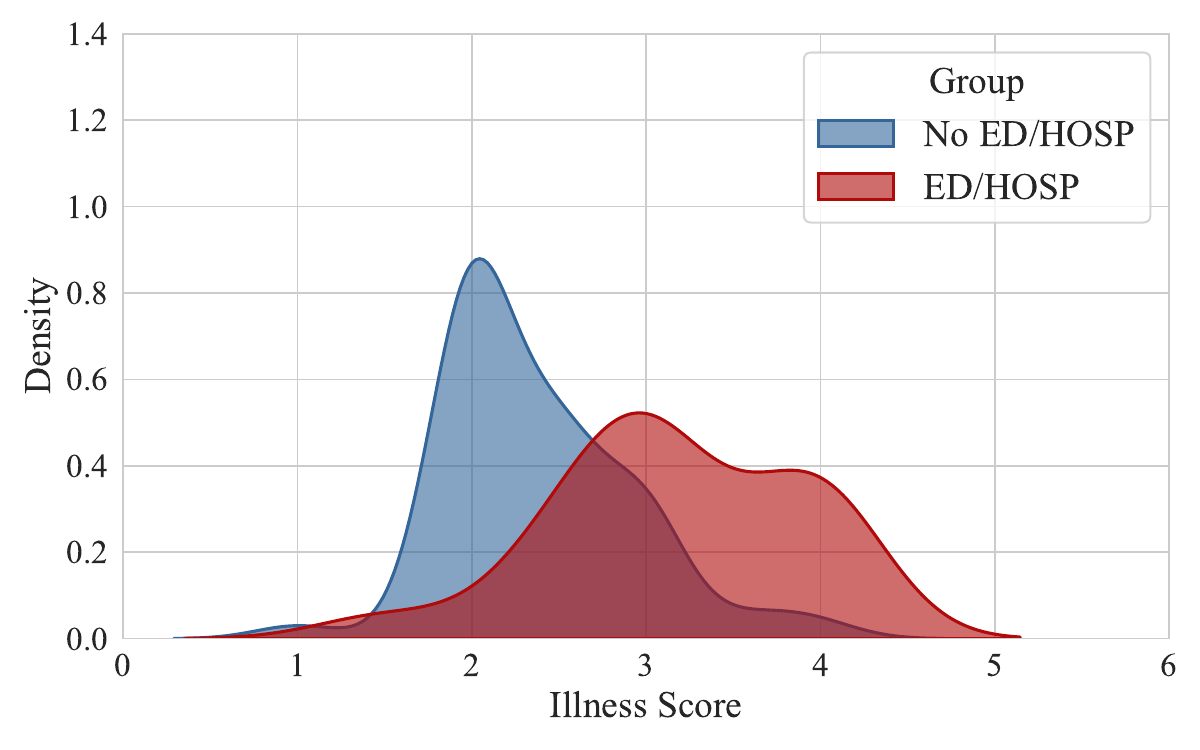}
        \caption{SOAP only}
        \label{fig:gpt41_kde_soap_only}
    \end{subfigure}
    \hfill
    \begin{subfigure}[t]{0.29\textwidth}
        \centering
        \includegraphics[width=\textwidth]{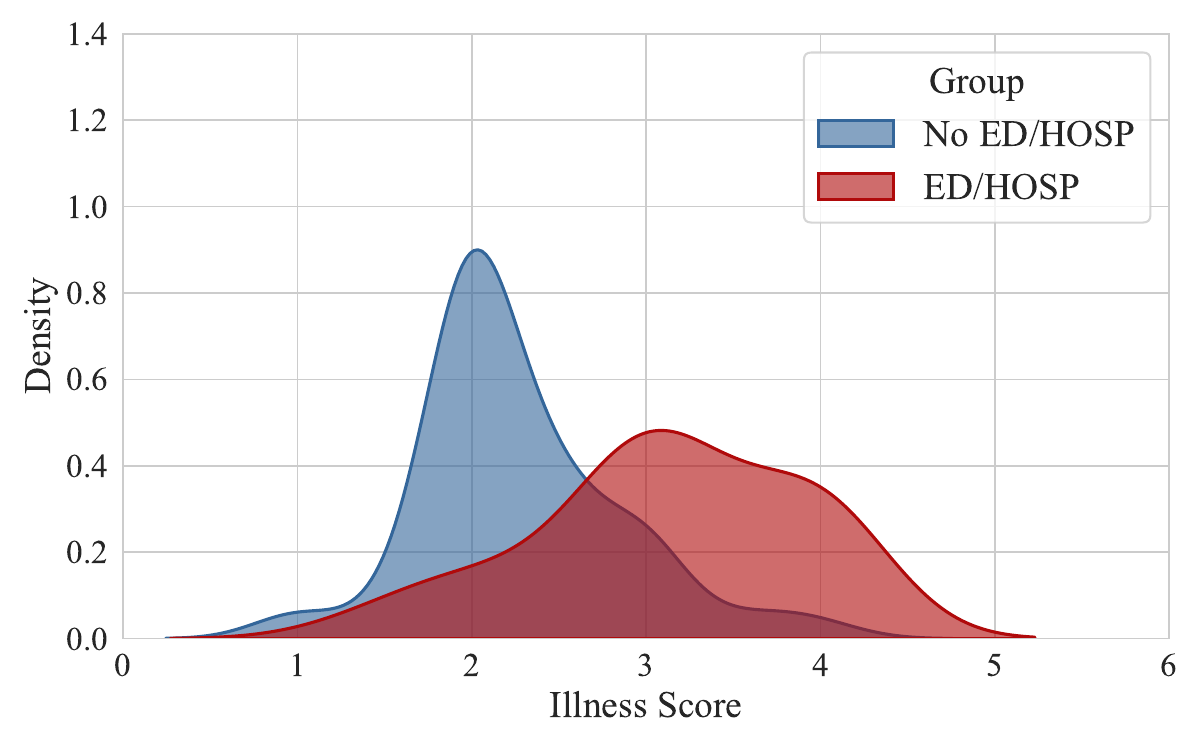}
        \caption{SOAP+VITAL}
        \label{fig:gpt41_kde_vitals_and_soap}
    \end{subfigure}
    
    \medskip
    \textbf{Qwen-3-8B}
    \smallskip
    
    \begin{subfigure}[t]{0.29\textwidth}
        \centering
        \includegraphics[width=\textwidth]{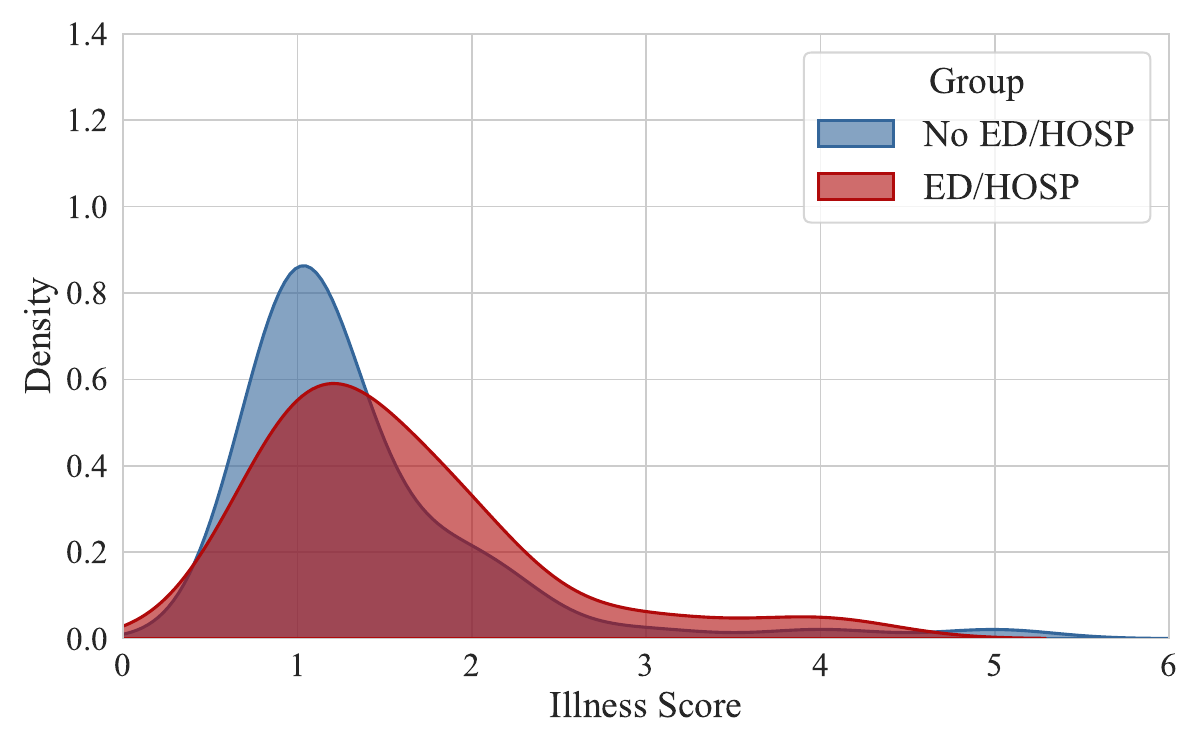}
        \caption{VITAL only}
        \label{fig:qwen3_8b_vitals}
    \end{subfigure}
    \hfill
    \begin{subfigure}[t]{0.29\textwidth}
        \centering
        \includegraphics[width=\textwidth]{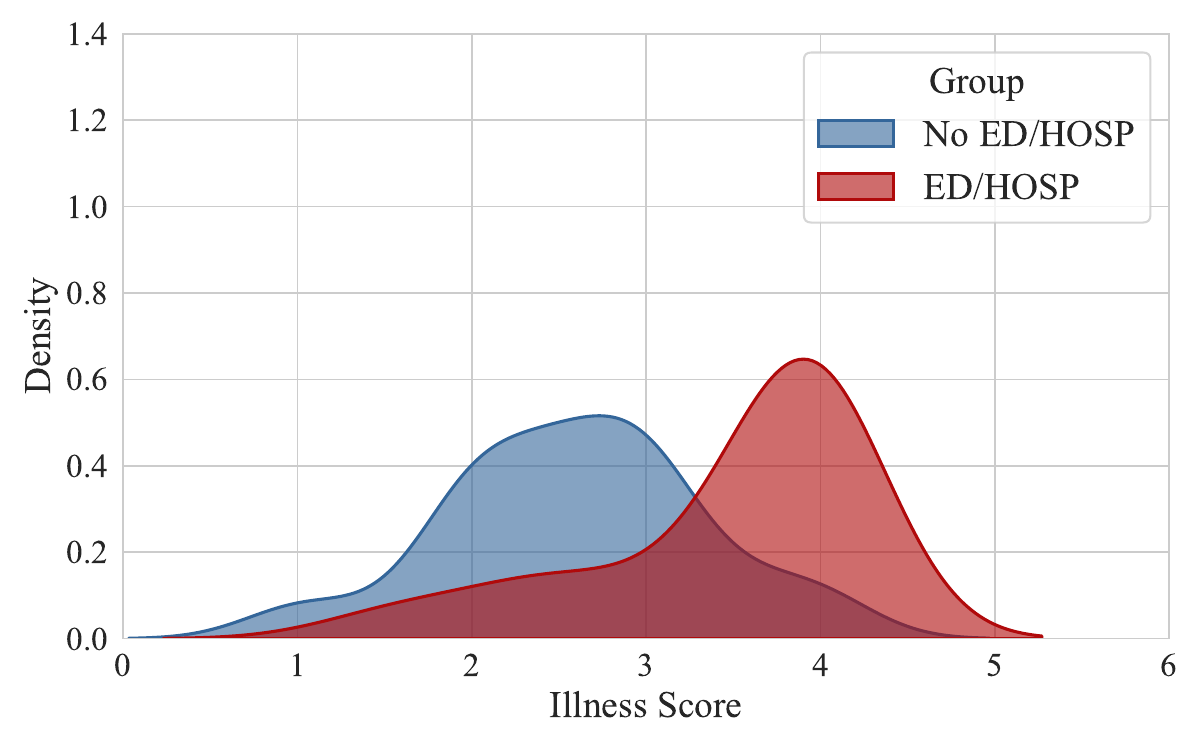}
        \caption{SOAP only}
        \label{fig:qwen3_8b_soap}
    \end{subfigure}
    \hfill
    \begin{subfigure}[t]{0.29\textwidth}
        \centering
        \includegraphics[width=\textwidth]{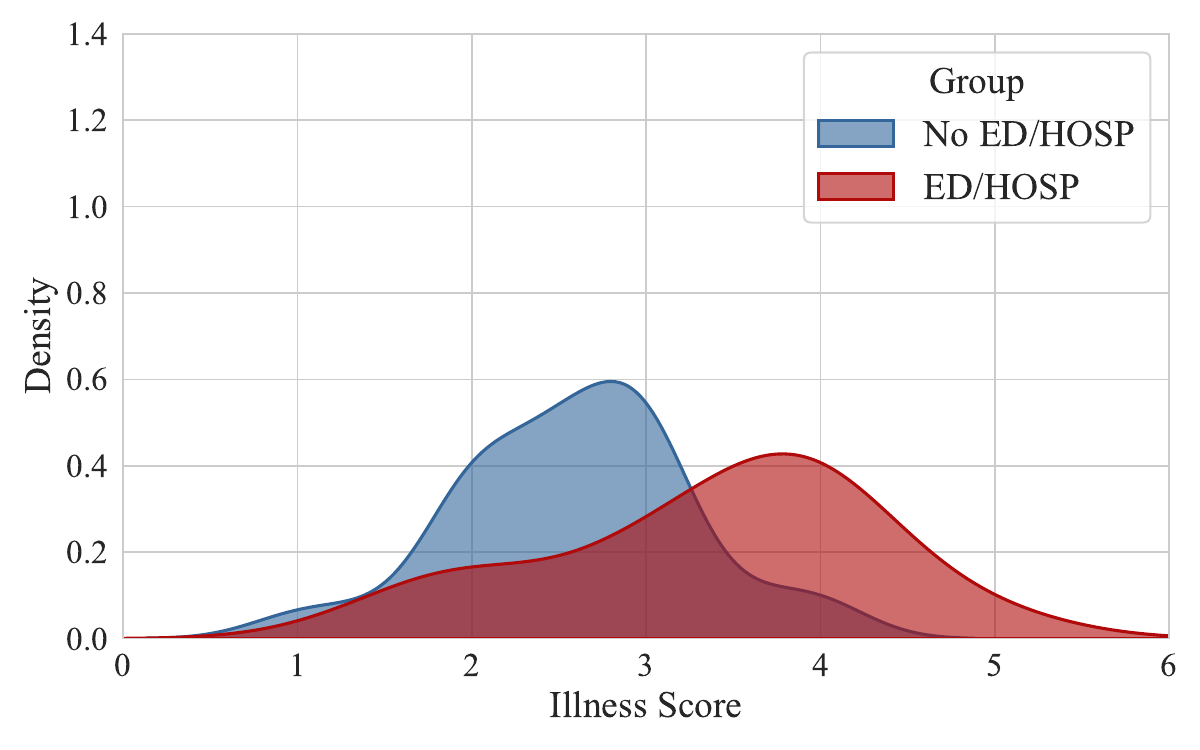}
        \caption{SOAP+VITAL}
        \label{fig:qwen3_8b_vitals_soap}
    \end{subfigure}
    
    \medskip
    
    \textbf{Llama-3.1-8B}
    \smallskip
    
    \begin{subfigure}[t]{0.29\textwidth}
        \centering
        \includegraphics[width=\textwidth]{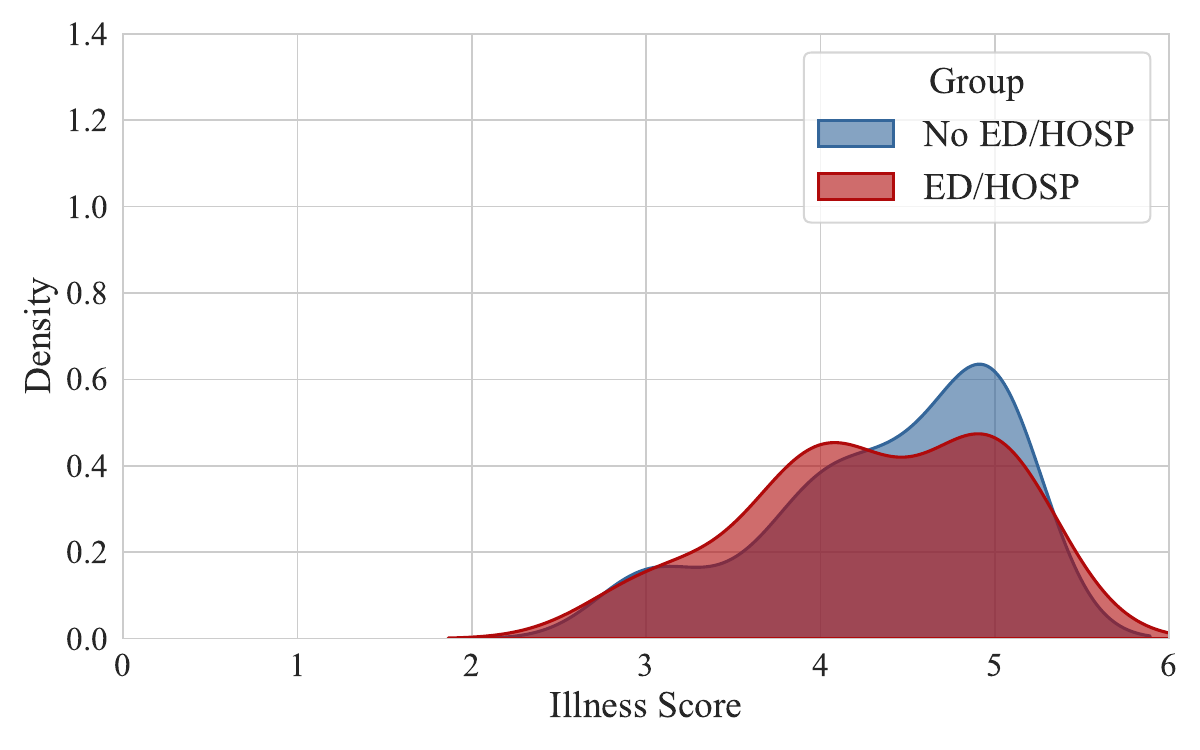}
        \caption{VITAL only}
        \label{fig:llama3_1_8b_vitals}
    \end{subfigure}
    \hfill
    \begin{subfigure}[t]{0.29\textwidth}
        \centering
        \includegraphics[width=\textwidth]{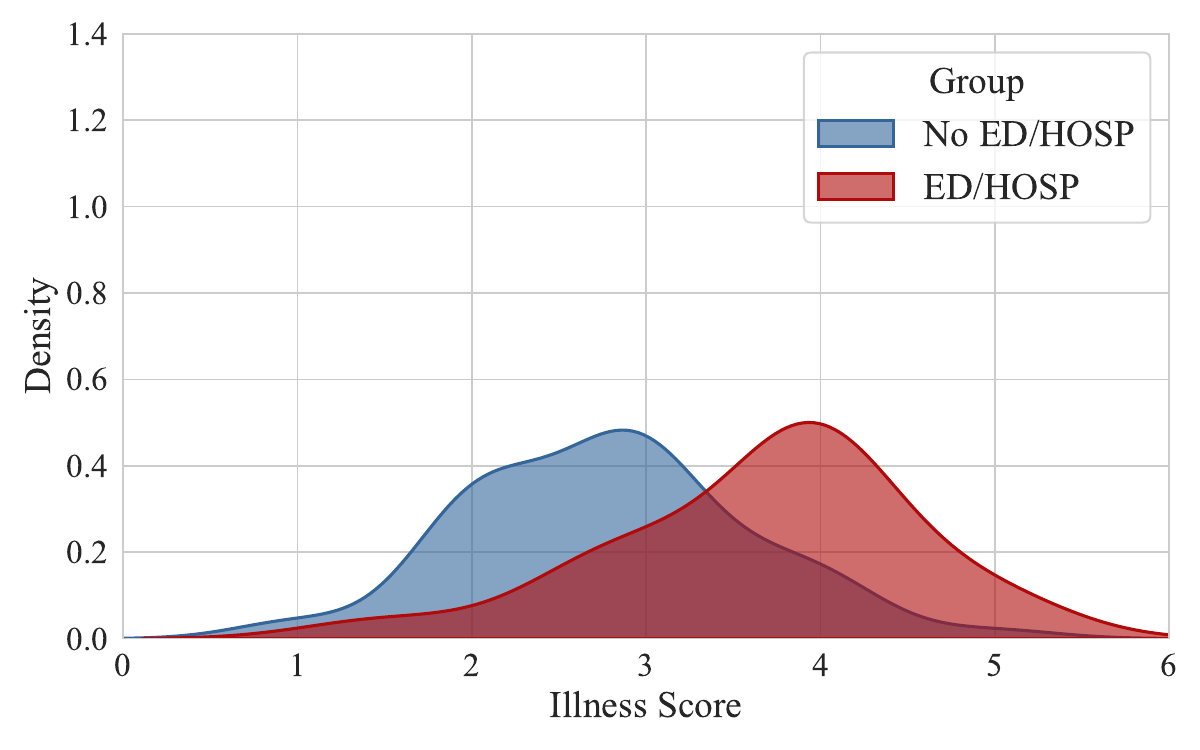}
        \caption{SOAP only}
        \label{fig:llama3_1_8b_soap}
    \end{subfigure}
    \hfill
    \begin{subfigure}[t]{0.29\textwidth}
        \centering
        \includegraphics[width=\textwidth]{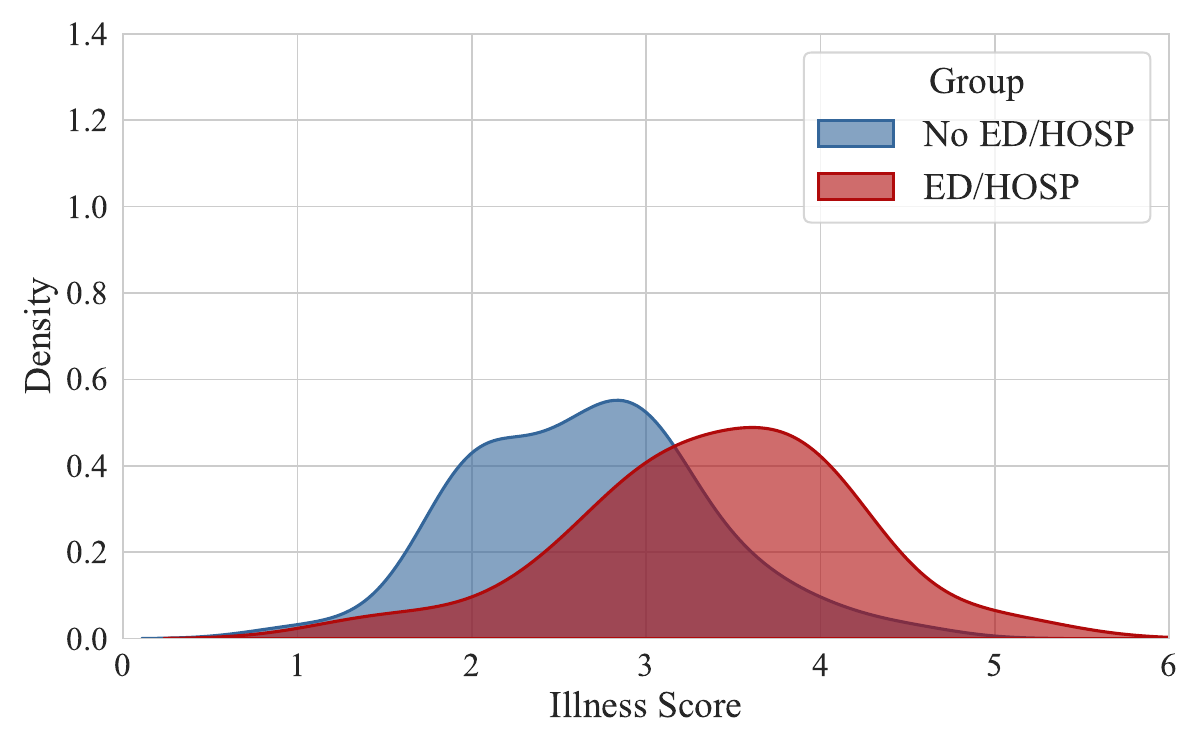}
        \caption{SOAP+VITAL}
        \label{fig:llama3_1_8b_vitals_soap}
    \end{subfigure}
    
    \medskip
    
    \textbf{MediPhi-Instruct}
    \smallskip
    
    \begin{subfigure}[t]{0.29\textwidth}
        \centering
        \includegraphics[width=\textwidth]{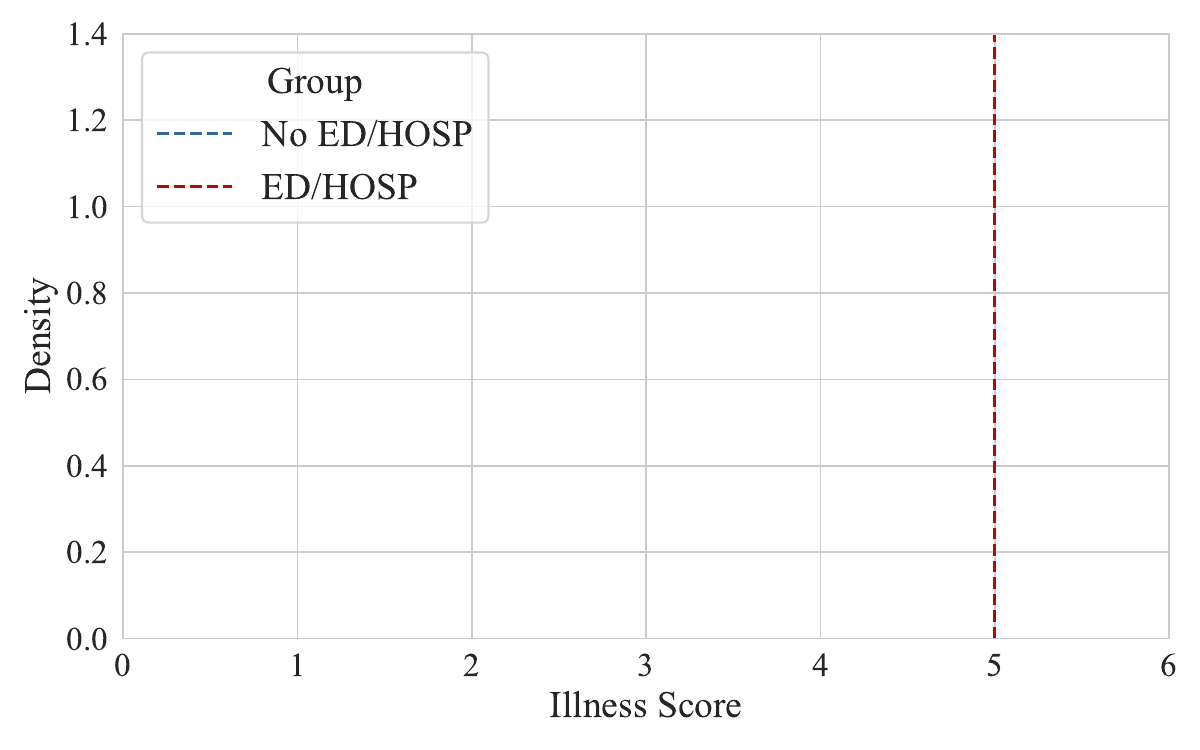}
        \caption{VITAL only}
        \label{fig:mediphi_vitals}
    \end{subfigure}
    \hfill
    \begin{subfigure}[t]{0.29\textwidth}
        \centering
        \includegraphics[width=\textwidth]{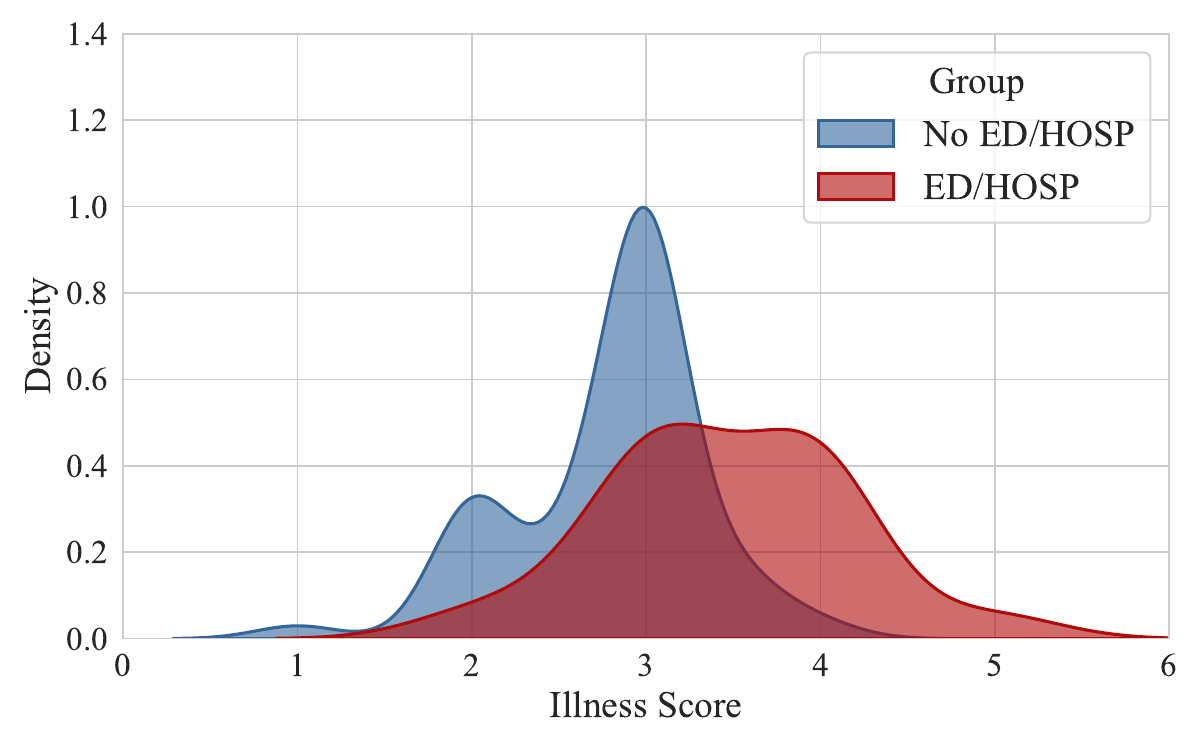}
        \caption{SOAP only}
        \label{fig:mediphi_soap}
    \end{subfigure}
    \hfill
    \begin{subfigure}[t]{0.29\textwidth}
        \centering
        \includegraphics[width=\textwidth]{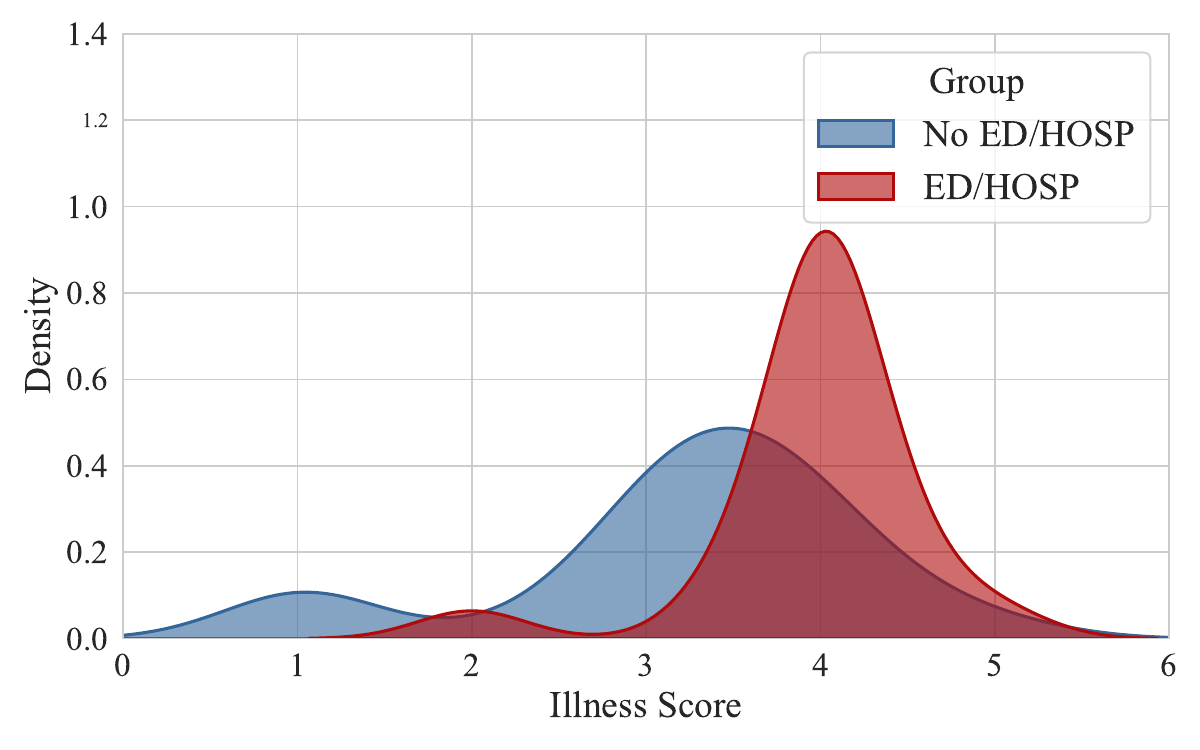}
        \caption{SOAP+VITAL}
        \label{fig:mediphi_vitals_soap}
    \end{subfigure}
    
    \caption{KDE plots of generated illness score distributions of No ED/HOSP and ED/HOSP groups. Each row shows results for a specific LLM, while columns represent different inputs. 
    Across most settings, ED/HOSP groups show higher average scores than No ED/HOSP groups, indicating alignment with clinical outcomes. SOAP-based inputs (SOAP only and SOAP+VITAL) show better group separation than VITAL only, with consistently higher scores for ED/HOSP patients.}
    \label{fig:all_4_models}
\end{figure}

The performance of four LLMs was evaluated: GPT-4.1, Qwen3-8B, Llama3.1-8B, and MediPhi-Instruct. For each LLM, we conducted an ablation study using different input types (VITAL only, SOAP only, and SOAP+VITAL) to determine which information an LLM can most effectively use to assess overall health. Figure~\ref{fig:all_4_models} shows Kernel Density Estimation (KDE) 
plots of resulting illness scores, with detailed summary statistics provided in the Appendix~\ref{sec:appendix_additional_results}. In all settings, ED/HOSP groups tend to have higher average illness scores compared to the No ED/HOSP group, except for MediPhi-Instruct and Llama3.1-8B with VITAL only. These results indicate that LLMs can effectively interpret SOAP notes, with some models also able to meaningfully incorporate vital signs, to generate scores aligning with the instruction that higher values correspond to more critical health conditions. We observe that illness scores derived from vital signs alone differ substantially from those generated with SOAP inputs (SOAP only and SOAP+VITAL). For VITAL only, the two groups tend to have similar distributions, reflecting that most patients’ vital signs fall within normal ranges. 
The similarity between SOAP only and SOAP+VITAL suggests that the LLM’s interpretation of SOAP notes is consistent, with vital signs providing subtle adjustments to the overall illness score. MediPhi-Instruct’s results deviate more from the other LLMs. With VITAL only inputs, it assigns a uniform score of 5 to all cases, requiring additional information to generate a meaningful score. 

\begin{figure}[htbp!]
    \centering
  \includegraphics[scale=0.83]{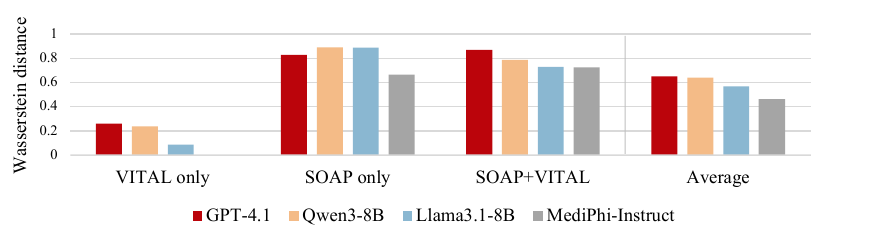} \hfill
  \caption{Wasserstein distances between illness score distributions of No ED/HOSP and ED/HOSP groups across four models. 
  A higher Wasserstein distance indicates better separation between groups. 
  SOAP notes provide substantially more discriminative information than vital signs alone. GPT-4.1 achieves the highest average performance and benefits from incorporating vital signs, while Qwen3-8B and Llama3.1-8B show degraded performance when vitals are added.
  }
    \label{fig:ED_histogram}
\end{figure}

To better compare the performance of each LLM, we calculate the Wasserstein distances (WD)~\citep{villani2009wasserstein} between the score distribution of the two groups (Figure~\ref{fig:ED_histogram}). Higher WD indicates greater separation between the two groups and is therefore considered better performance. The results show that SOAP notes provide substantially more information for overall health assessment than vital signs, as VITAL only performed much lower across all LLMs. In addition, GPT-4.1 and MediPhi-Instruct benefit from including vital signs, with SOAP+VITAL inputs outperforming SOAP-only inputs. In contrast, Qwen3-8B and Llama3.1-8B perform worse when vital inputs are added, indicating that they cannot effectively integrate this information. 
Given that GPT-4.1 demonstrates the best performance on average and greater robustness with or without vital information, we selected its SOAP+VITAL results for the subsequent acoustic analysis. 

\subsection{ALM for Speaker Age and Gender Classification}

\begin{figure}[htbp!]
    \centering
  \includegraphics[scale=0.9]{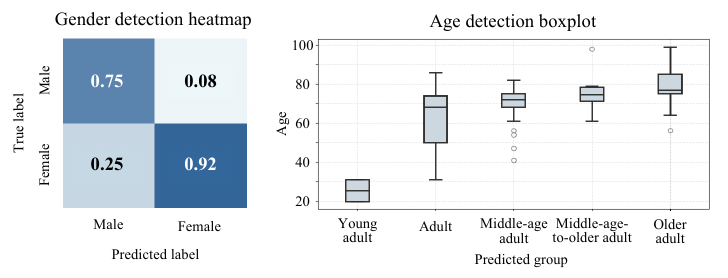} \hfill
  \caption{GPT-4o-audio performance on a baseline speaker classification tasks. 
\textbf{Left}: Confusion matrix showing GPT-4o-audio achieves 0.8625 accuracy on gender detection.
\textbf{Right}: Age prediction boxplot showing predicted age categories that align well with true ages. 
}
    \label{fig:age_gender}
\end{figure}

Figure~\ref{fig:age_gender} presents the results for GPT-4o-audio. For gender prediction, the model achieves an accuracy of 0.8625, with higher precision for female voices than for male. For age prediction, the model classifies speakers into categories: young adult, adult, middle-aged, middle-aged to older, and older adult. The predicted age categories largely correspond to the speakers’ true ages. These results indicate that GPT-4o-audio can reliably capture speaker characteristics from audio signals, supporting more complex acoustic analyses, such as health assessment. Compared with GPT-4o-audio, GPT-4o-mini-audio shows a reduced ability in vocal biomarker extraction, e.g., achieving only 0.7484 gender prediction accuracy. Furthermore, the open-source Qwen2-Audio-7B-Instruct demonstrates intermediate performance, with an accuracy of 0.8064 on gender prediction.
Given its stronger performance in extracting basic speaker traits, the following ALM analyses are based on GPT-4o-audio, while results for GPT-4o-mini and Qwen2-Audio-7B-Instruct are provided in Appendix~\ref{sec:gpt-4o-mini}.

\subsection{ALM Analysis of Clinician and Patient Voices}
\begin{figure}[htbp!]
    \centering
  \includegraphics[scale=0.71]{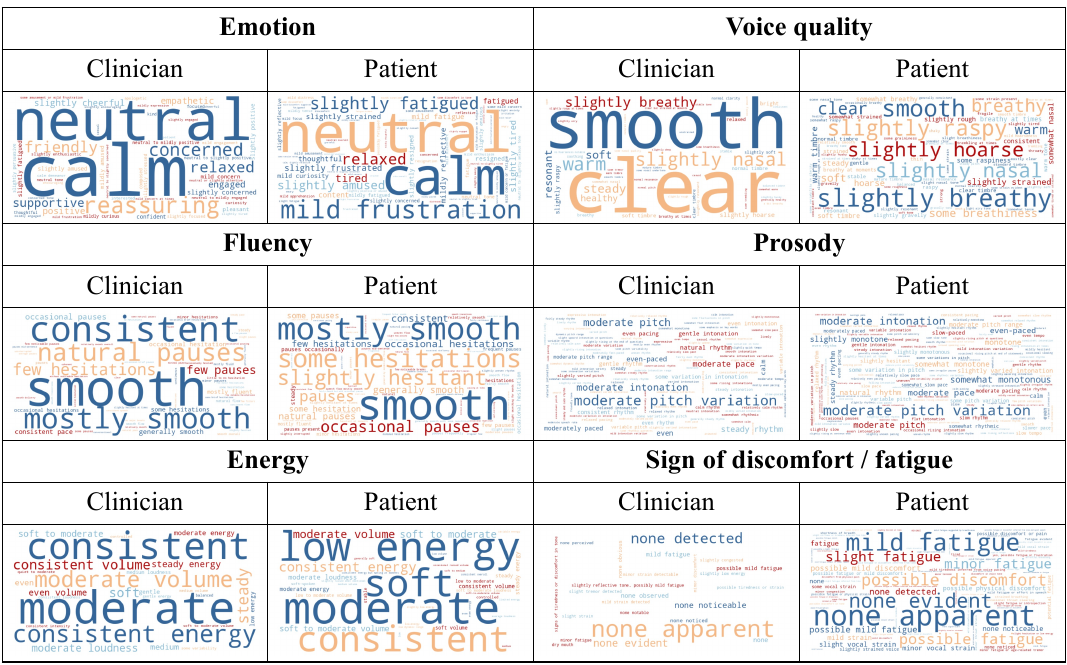} \hfill
  \caption {Wordcloud of GPT-4o-audio generated phrase frequencies for clinician and patient voices.}
    \label{fig:world_cloud}
\end{figure}

Figure~\ref{fig:world_cloud} presents wordclouds of ALM-generated phrases describing various aspects of the speech signals from each group. The results show that the most commonly identified emotions for both clinicians and patients were neutral and calm in home care conversations. However, clinicians were more frequently described using positive emotions whereas patients’ speech was more often associated with negative emotions like frustration. When examining voice quality, clinicians’ voices were typically characterized as smooth and clear, consistent with professional speech patterns. Conversely, patients’ voices were more likely to be described as raspy, breathy, or nasal, which may be indicative of underlying health conditions affecting their vocal quality. For pronunciation, both groups showed moderate pitch variation in prosody, but the patients’ speech tended to be more monotone. For fluency, clinicians generally spoke in a consistent and smooth manner and patients tended to hesitate more often. For energy level, clinician speech is most often described as consistent or moderate. While these descriptors are also common for patients, their speech is also frequently characterized as low energy and soft. Lastly, indicators of discomfort and fatigue were more prominent in patients’ speech. Overall, the ALM’s descriptions of clinician voices align more closely with characteristics typically associated with healthier individuals, whereas descriptions of patient voices reflect traits indicative of poorer health. These results demonstrate that the ALM-generated vocal descriptions are interpretable and capable of distinguishing individuals with differing health conditions based on their speech signals.

\subsection{ALM Analysis of Patient Voices Across Health Conditions}

\begin{figure}[htbp!]
    \centering
  \includegraphics[scale=0.79]{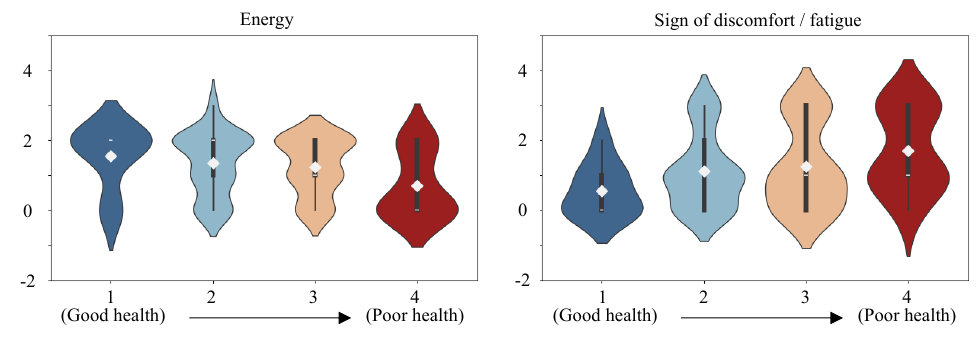} \hfill
  \caption{Violin plots showing the relationship between GPT-4o-audio extracted acoustic features and GPT-4.1 generated illness scores. Diamond markers indicate group means, and the width represents the density of observations at that value. 
  \textbf{Left}: Energy levels decline as health worsens. 
    \textbf{Right}: Signs of discomfort/fatigue increase with declining health status.
  }
    \label{fig:violin_acoustic}
\end{figure}

We used illness scores generated by the GPT-4.1 (SOAP+VITAL) to group samples accordingly and analyze the corresponding changes in the acoustic descriptions. Table~\ref{tab:top_3_important_phrases} presents the top three important phrases identified using Term Frequency–Inverse Document Frequency (TF-IDF) for each illness score group. The observed trend is consistent with the clinician–patient comparison discussed above. Across all groups, emotional tone is predominantly neutral or calm. As health status declines from relatively healthy to worse, voice quality tends to shift from smooth to increasingly breathy, raspy, or hoarse. Fluency changes from smoother delivery to more hesitant speech, while prosody changes from moderate pitch variation to a more monotone pattern.
\begin{table}[t]
\centering
\caption{Top-3 acoustic descriptions for patients per health condition group using TF-IDF analysis. The phrases within each group are ranked from most to least frequently generated by GPT-4o-audio.}
\label{tab:top_3_important_phrases}
\resizebox{\textwidth}{!}{%
\begin{tabular}{l|l|l|l|l}
\hline
\hline
\textbf{Acoustic}
& \multicolumn{4}{c}{\textbf{Illness Score}} \\
\cline{2-5}
\textbf{Feature} & 
\begin{tabular}[t]{@{}c@{}}
\textbf{1} \textbf{(Good health)}
\end{tabular} & 
\begin{tabular}[t]{@{}c@{}}
\textbf{2}
\end{tabular} & 
\begin{tabular}[t]{@{}c@{}}
\textbf{3}
\end{tabular} & 
\begin{tabular}[t]{@{}c@{}}
\textbf{4} \textbf{(Poor health)}
\end{tabular} \\
\hline
\textbf{Emotion} & 
\begin{tabular}[t]{@{}l@{}}
Neutral \\ Calm \\ Some mild concern
\end{tabular} & 
\begin{tabular}[t]{@{}l@{}}
Neutral \\ Calm \\ Slightly fatigued
\end{tabular} & 
\begin{tabular}[t]{@{}l@{}}
Neutral \\ Calm \\ Mild frustration
\end{tabular} & 
\begin{tabular}[t]{@{}l@{}}
Neutral \\ Calm \\ Resigned
\end{tabular} \\
\hline
\begin{tabular}[t]{@{}l@{}}
\textbf{Voice} \\ \textbf{Quality}
\end{tabular} &  
\begin{tabular}[t]{@{}l@{}}
Smooth \\ Soft \\ Slightly breathy
\end{tabular} & 
\begin{tabular}[t]{@{}l@{}}
Smooth \\ Slightly breathy \\ Slightly raspy
\end{tabular} & 
\begin{tabular}[t]{@{}l@{}}
Slightly breathy \\ Smooth \\ Slightly nasal
\end{tabular} & 
\begin{tabular}[t]{@{}l@{}}
Slightly breathy \\ Slightly raspy \\ Slightly hoarse
\end{tabular} \\
\hline
\textbf{Prosody} & 
\begin{tabular}[t]{@{}l@{}}
Moderate pitch variation \\ Even \\ Steady rhythm
\end{tabular} & 
\begin{tabular}[t]{@{}l@{}}
Moderate intonation \\ Moderate pitch variation \\ Moderate pace
\end{tabular} & 
\begin{tabular}[t]{@{}l@{}}
Moderate pitch variation \\ Paced \\ Moderate intonation
\end{tabular} & 
\begin{tabular}[t]{@{}l@{}}
Somewhat monotonous \\ Monotone \\ Moderate pitch variation
\end{tabular} \\
\hline
\textbf{Fluency} & 
\begin{tabular}[t]{@{}l@{}}
Mostly smooth \\ Smooth \\ Occasional pauses
\end{tabular} & 
\begin{tabular}[t]{@{}l@{}}
Smooth \\ Mostly smooth \\ Occasional pauses
\end{tabular} & 
\begin{tabular}[t]{@{}l@{}}
Occasional hesitations \\ Slightly hesitant \\ Mostly smooth
\end{tabular} & 
\begin{tabular}[t]{@{}l@{}}
Slightly hesitant \\ Some hesitations \\ Pauses frequently
\end{tabular} \\
\hline
\hline
\end{tabular}
}
\end{table}
For features with a continuous scale, such as energy level and potential signs of fatigue or discomfort, we converted the descriptive phrases into numerical values to enable quantitative analysis. For energy, we mapped the output description phrases to pre-defined categories by calculating sentence similarity~\citep{bge-m3} between the original phrase and the following reference categories: \emph{low/soft}, \emph{soft to moderate}, \emph{moderate,} and \emph{consistent/steady}, corresponding to energy levels 1 to 4. For example, phrases most similar to \emph{low/soft} were assigned an energy score of 1, whereas those most similar to \emph{consistent/steady} received a score of 4. Similarly, for possible signs of discomfort and fatigue, we defined anchor categories as follows: \emph{no evident/no apparent}, \emph{mild discomfort/fatigue}, \emph{moderate discomfort/fatigue}, and \emph{noticeable discomfort/fatigue}. Output phrases were assigned numeric scores based on sentence similarity to these anchors, allowing quantification of severity on a scale from 1 (no discomfort/fatigue) to 4 (noticeable discomfort/fatigue). Results shown in Figure~\ref{fig:violin_acoustic} show that both lower energy levels and increased signs of fatigue or discomfort detected by the ALM are indicative of poorer health status. 

Overall, our results suggest that speech signals can exhibit distinct patterns associated with differences in health conditions. The ALM extracts interpretable acoustic descriptions that help characterize these patterns. While acoustic signals alone should not be used for diagnosis, such vocal characteristics could serve as supplementary cues in healthcare monitoring systems.

\section{Limitations}
Our study analyzes audio from a real-world home care setting, which presents challenges such as background noise, variable recording conditions, and frequent overlapping speech. While our automated acoustic processing pipeline substantially mitigates their impact on audio analysis, it remains imperfect and occasionally fails to fully isolate speakers, resulting in leakage from others and potentially blurring speaker-specific signals. Also, we examine only the first 30 seconds of the processed patient speech for acoustic analysis. This may omit important acoustic cues that occur later, including changes in the speaker’s health condition.

Although we explicitly instruct the ALM to avoid transcribing the audio and to disregard linguistic content, it may not truly adhere to this, relying on contextual information to support its outputs. Further work is needed to assess the degree to which the ALM can disentangle acoustic from linguistic information. In addition, GPT-Audio declines to provide direct clinical judgments based solely on acoustic signals, while accepting the analysis of various acoustic features. The choice of prompt influences the ALM’s outputs, but this effect was not examined in depth in the present study.

Moreover, this study focuses on analyzing acoustic signals in relation to overall health conditions, but in real-world healthcare contexts, voice alone should not serve as the sole basis for diagnosis. Instead, it should be regarded as a complementary signal, useful for capturing conditions not reflected in textual or numerical records, or in situations where such records are incomplete or inconsistent.

Finally, although experimental results demonstrate that our proposed LLM-generated illness scores align with clinical outcomes, we did not compare these scores directly with clinician annotations. This is because clinical judgments for overall health assessment are highly subjective and require extensive expertise, making such annotation both challenging and costly. Given the complexity and resource demands of this task, comparing LLM labels with clinician labels is beyond the scope of the present study.

\section{Conclusion}
This study investigates the use of LLMs to automatically assess patient health by generating holistic illness scores in a home care setting. We find that LLMs can effectively interpret SOAP notes extracted from visit transcripts to produce scores that align with clinical outcomes, whereas vital signs are less informative. To our knowledge, this is the first work to apply ALMs to assess patient health directly from speech, showing that they can capture clinically relevant acoustic patterns that, when quantified, correlate strongly with our LLM-based illness scores. Furthermore, we demonstrate ALMs' ability to describe acoustic features in interpretable plain language, enabling clinicians to incorporate speech as a meaningful health biomarker, adding another dimension of patient monitoring. Although our approach faces several challenges, it demonstrates the potential of leveraging generative AI for multimodal health assessment in real-world care and highlights the need to further explore the opportunities that ALMs offer for healthcare applications.

\section*{Acknowledgments}
This work was supported by the National Institutes of Health under Grant No. 1R01AG081928-01, in addition to Grant No. P30AG073105 and R00AG076808 under the National Institute of Aging.

\bibliographystyle{plainnat}
\bibliography{bibliography}


\appendix
\newpage

\section{Prompts}\label{sec:appendix_prompts}

\begin{figure}[htbp!]
    \centering
  \includegraphics[width=\linewidth]{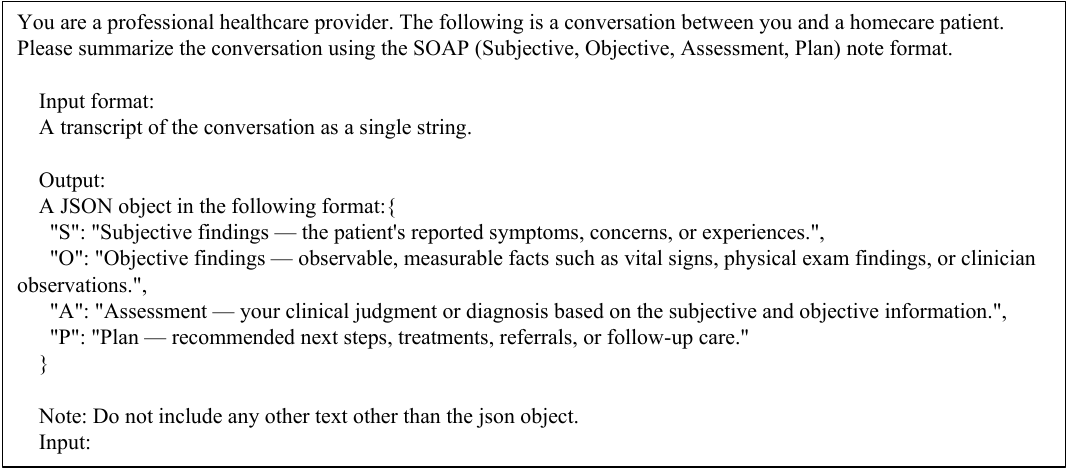} \hfill
  \caption {Prompt template used for LLM-based SOAP note summarization from conversation transcripts.}
    \label{fig:prompt_soap}
\end{figure}

\textbf{Prompt for generating SOAP note} (Figure~\ref{fig:prompt_soap})

\begin{figure}[htbp!]
    \centering
  \includegraphics[width=\linewidth]{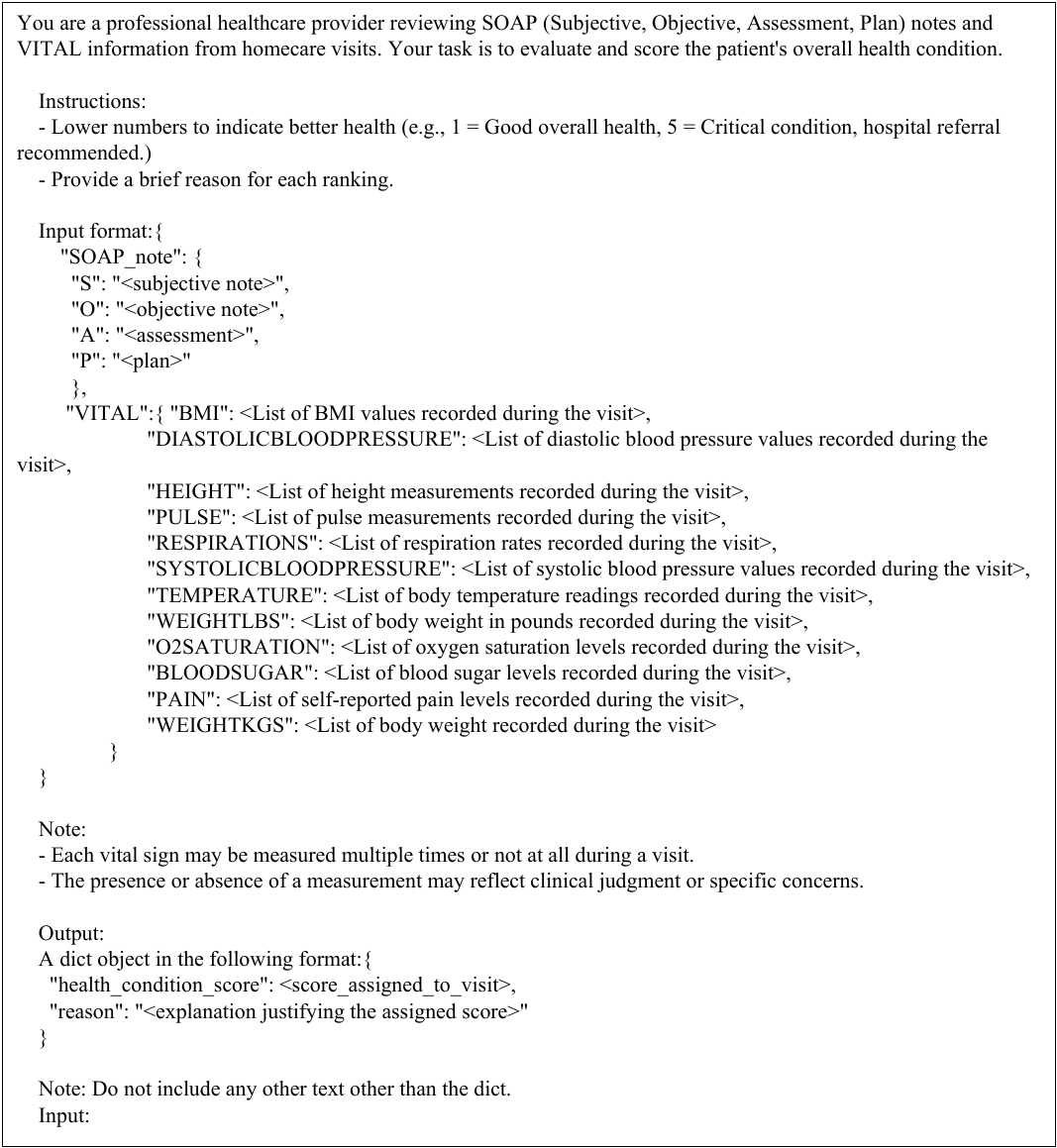} \hfill
  \caption {Prompt template used for LLM-based illness scoring from SOAP notes and vital signs.}
    \label{fig:prompt_illness}
\end{figure}

\textbf{Prompt for LLM-based holistic illness severity scoring} (Figure~\ref{fig:prompt_illness})

\begin{figure}[htbp!]
    \centering
  \includegraphics[width=0.9\linewidth]{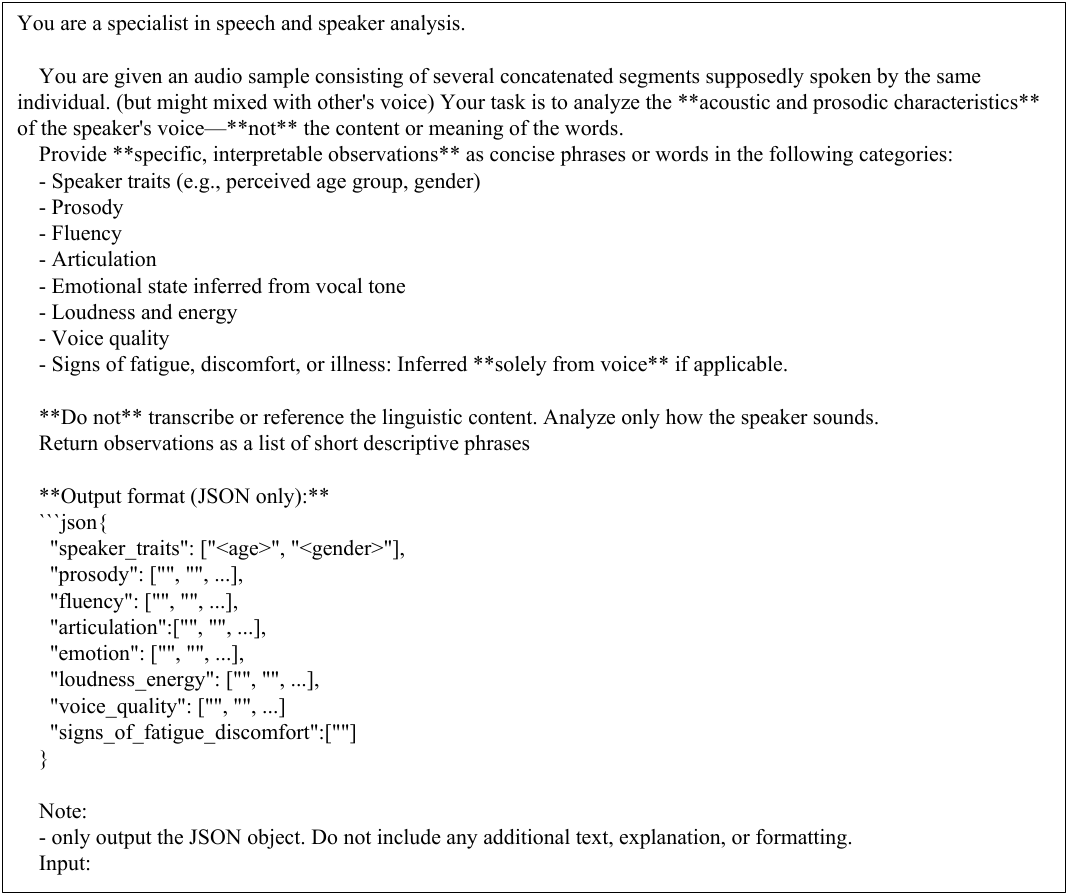} \hfill
  \caption {Prompt template used for ALM acoustic analysis instructing the model to analyze acoustic characteristics without transcribing linguistic content.}
    \label{fig:prompt_alm}
\end{figure}

\textbf{Prompt for ALM acoustic analysis} (Figure~\ref{fig:prompt_alm}): We ask the ALM to generate a list of phrases rather than full sentences or fixed categories. This approach is motivated by several considerations: full sentences are difficult to evaluate because the same description can be expressed in many different ways, even though the underlying meaning is the same. Fixed categories, on the other hand, would constrain the model’s output.

\newpage

\section{Additional Results}

\subsection{Summary statistics for LLM-based holistic illness severity scoring (Table~\ref{tab:illness_scores})}\label{sec:appendix_additional_results} 

\begin{table}[hbpt!]
\centering
\caption{Summary statistics for LLM-generated illness scores across different inputs types.}
\label{tab:illness_scores}
\begin{tabular}{llccccccc}
\hline
\hline
{\textbf{Model}} & {\textbf{Input}} & \multicolumn{3}{c}{\textbf{No ED/HOSP}} & \multicolumn{3}{c}{\textbf{ED/HOSP}} \\
\cmidrule(lr){3-5} \cmidrule(lr){6-8}
& & \textbf{Min} & \textbf{Max} & \textbf{Mean} & \textbf{Min} & \textbf{Max} & \textbf{Mean} \\
\midrule
{GPT-4.1} 
& VITAL only & 1.0 & 4.0 & 1.60 & 1.0 & 2.0 & 1.73 \\
& SOAP only & 1.0 & 4.0 & 2.38 & 1.5 & 4.0 & 3.18 \\
& SOAP+VITAL & 1.0 & 4.0 & 2.27 & 1.5 & 4.0 & 3.14 \\
\midrule
{Llama3.1-8B} 
& VITAL only & 3.0 & 5.0 & 4.38 & 3.0 & 5.0 & 4.30 \\
& SOAP only & 1.0 & 5.0 & 2.78 & 1.5 & 5.0 & 3.67 \\
& SOAP+VITAL & 1.0 & 4.5 & 2.67 & 1.5 & 5.0 & 3.39 \\
\midrule
{MediPhi-Instruct} 
& VITAL only & 5.0 & 5.0 & 5.00 & 5.0 & 5.0 & 5.00 \\
& SOAP only & 1.0 & 4.0 & 2.79 & 2.0 & 5.0 & 3.46 \\
& SOAP+VITAL & 1.0 & 5.0 & 3.25 & 2.0 & 5.0 & 3.97 \\
\midrule
{Qwen3-8B} 
& VITAL only & 1.0 & 5.0 & 1.37 & 1.0 & 4.0 & 1.57 \\
& SOAP only & 1.0 & 4.0 & 2.60 & 1.5 & 4.0 & 3.49 \\
& SOAP+VITAL & 1.0 & 4.0 & 2.59 & 1.5 & 5.0 & 3.38 \\
\hline
\hline
\end{tabular}
\end{table}

\subsection{Examples of LLM’s rationale for illness scores}

\textbf{1 (Good health)}: The patient demonstrates excellent recovery after the procedure, with normal vital signs, good functional mobility, independence with daily activities and exercise, and no current pain or acute issues. Minor symptoms (occasional dizziness, mild intermittent pain) are self-managed and non-limiting. Discharge from home care is appropriate; following up with specialists is planned for further optimization. 

\textbf{2}: Patient is making good post-operative progress with stable vital signs (BP 120/80, pulse 69, respirations 13, temp 98.1). Incision is healing well with no signs of infection, pain is mild and well controlled, and patient is able to perform activities of daily living with minimal assistance. There are no reports of serious symptoms or complications. Continued monitoring and gradual reduction in opioid use is appropriate for this recovery stage. Only mild issues such as minor discomfort and adhesive residue are present.

\textbf{3}: The patient is stable with no acute distress, pain, or new skin issues, and displays improved mobility and motivation. However, there is persistent right-sided paralysis with significant ongoing mobility limitations, elevated blood glucose consistently above target indicating suboptimal diabetes control, and safety concerns regarding stairs. Edema is improving and vital signs are otherwise stable. The condition is chronic and requires ongoing management but does not meet criteria for hospital referral at this time.

\textbf{4 (Poor health)}: Patient is elderly, recently hospitalized, and requires ongoing oxygen therapy for chronic respiratory issues with worsening dyspnea. The patient is unable to stand or transfer independently and needs significant physical assistance and equipment for mobility. There is concern for hypoxemia requiring further respiratory management (possible BIPAP), and care is complicated by coordination of multiple services and progressive de-conditioning. Although vitals are stable, the overall functional and respiratory status is poor, indicating high risk but not immediate critical instability.

\subsection{Analysis of articulation}

\begin{figure}[htbp!]
    \centering
  \includegraphics[scale=1.0]{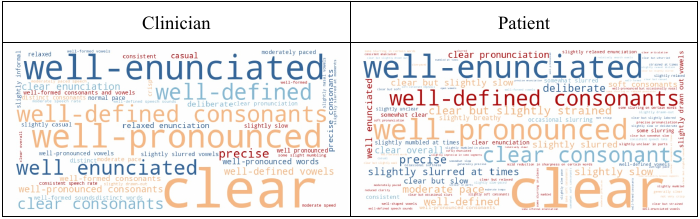} \hfill
  \caption {Wordcloud of GPT-4o-audio generated phrase frequencies describing articulation in clinician and patient voices.}
    \label{fig:world_cloud_articulation}
\end{figure}

Figure~\ref{fig:world_cloud_articulation} and Table~\ref{tab:top_3_important_phreases} present the ALM analysis results for articulation with GPT-4o-audio. In home care visits, articulation is a less informative indicator of health condition, as participants generally speak in their familiar language and therefore articulate their words clearly.

\begin{table}[hbpt!]
\centering
\caption{Top-3 articulation descriptions for patients per health condition group using TF-IDF analysis. The phrases within each group are ranked from most to least frequently generated by GPT-4o-audio.}
\label{tab:top_3_important_phreases}
\begin{tabular}{l|l|l|l|l}
\hline
\hline
\textbf{Acoustic}
& \multicolumn{4}{c}{\textbf{Illness Score}} \\
\cline{2-5}
\textbf{Feature} & 
\begin{tabular}[t]{@{}c@{}}
\textbf{1} \textbf{(Good health)}
\end{tabular} & 
\begin{tabular}[t]{@{}c@{}}
\textbf{2}
\end{tabular} & 
\begin{tabular}[t]{@{}c@{}}
\textbf{3}
\end{tabular} & 
\begin{tabular}[t]{@{}c@{}}
\textbf{4} \textbf{(Poor health)}
\end{tabular} \\
\hline
\textbf{Articulation} & 
\begin{tabular}[t]{@{}l@{}}
Clear \\ Well \\ Pronounced
\end{tabular} & 
\begin{tabular}[t]{@{}l@{}}
Clear \\ Well \\ Pronounced
\end{tabular} & 
\begin{tabular}[t]{@{}l@{}}
Clear \\ Well \\ Enunciated
\end{tabular} & 
\begin{tabular}[t]{@{}l@{}}
Clear \\ Slightly unclear \\ Deliberate
\end{tabular} \\
\hline
\hline
\end{tabular}
\end{table}

\section{Experimental results of GPT-4o-mini-audio and Qwen2-Audio-7B-Instruct}\label{sec:gpt-4o-mini}

\begin{figure}[htbp!]
    \centering
  \includegraphics[scale=1.1]{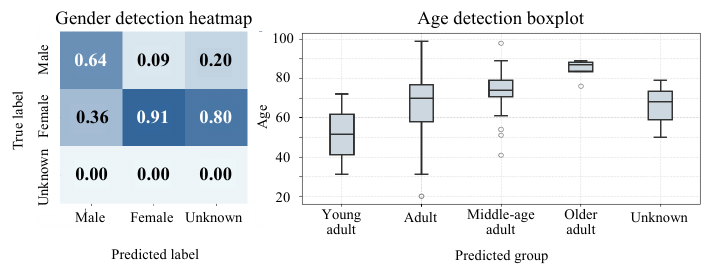} \hfill
  \caption {GPT-4o-mini performance in gender and age classification.}
    \label{fig:4omini-gender-age}
\end{figure}

\subsection{Age and Gender Classification with GPT-4o-mini-audio (Figure~\ref{fig:4omini-gender-age})}

\begin{figure}[htbp!]
    \centering
  \includegraphics[scale=1.1]{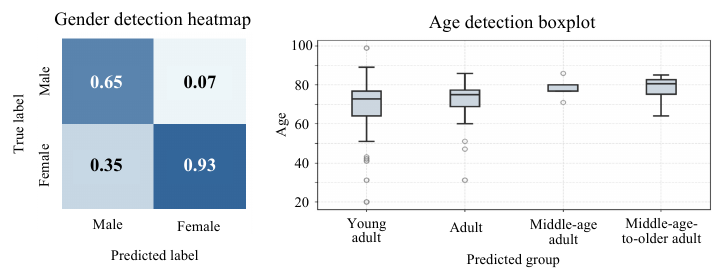} \hfill
  \caption {Qwen2-Audio-7B-Instruct performance in gender and age classification.}
    \label{fig:Qwen2-Audio-7B-Instructi-gender-age}
\end{figure}

\subsection{Age and Gender Classification with Qwen2-Audio-7B-Instruct (Figure~\ref{fig:Qwen2-Audio-7B-Instructi-gender-age})}

\newpage

\begin{figure}[htbp!]
    \centering
  \includegraphics[scale=0.73]{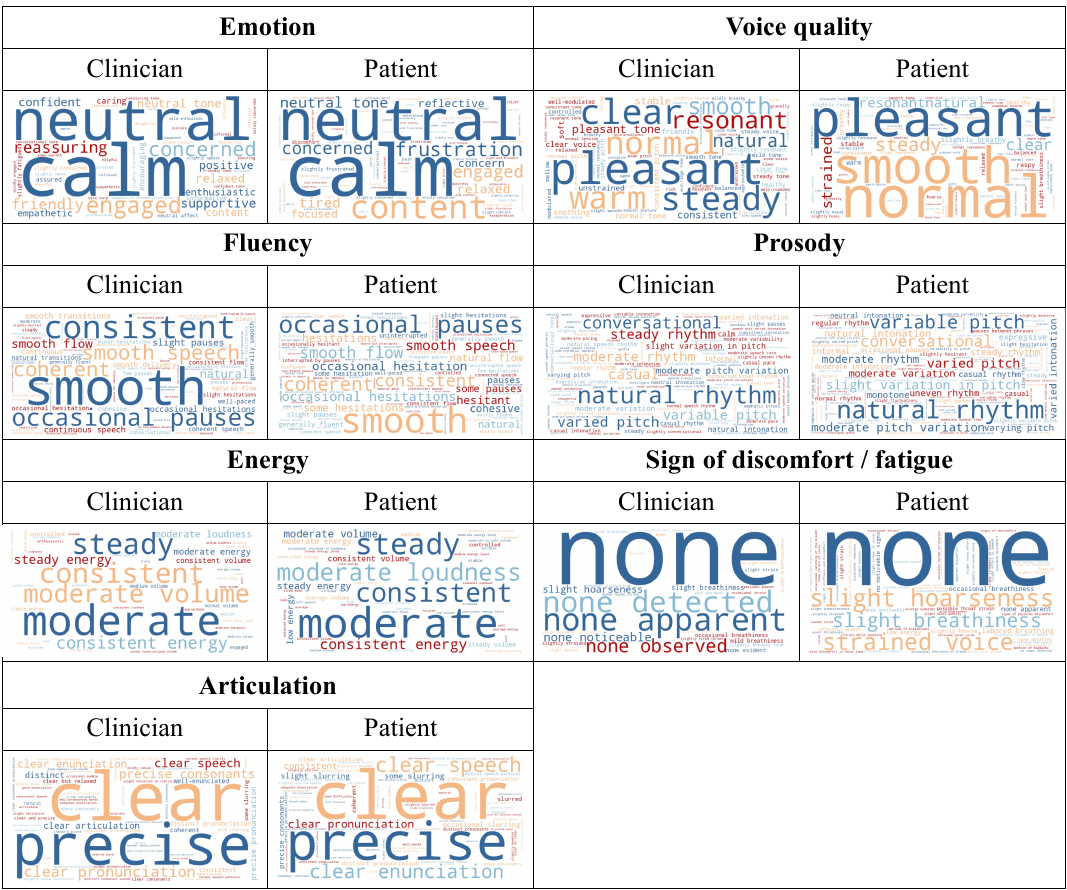} \hfill
  \caption {Wordcloud of GPT-4o-mini-audio-generated phrase frequencies for clinician and patient voices.}
    \label{fig:4omini-wordcloud}
\end{figure}

\subsection{Analysis of Clinician and Patient Voices with GPT-4o-mini-audio (Figure~\ref{fig:4omini-wordcloud})}

\newpage

\begin{figure}[htbp!]
    \centering
  \includegraphics[scale=0.73]{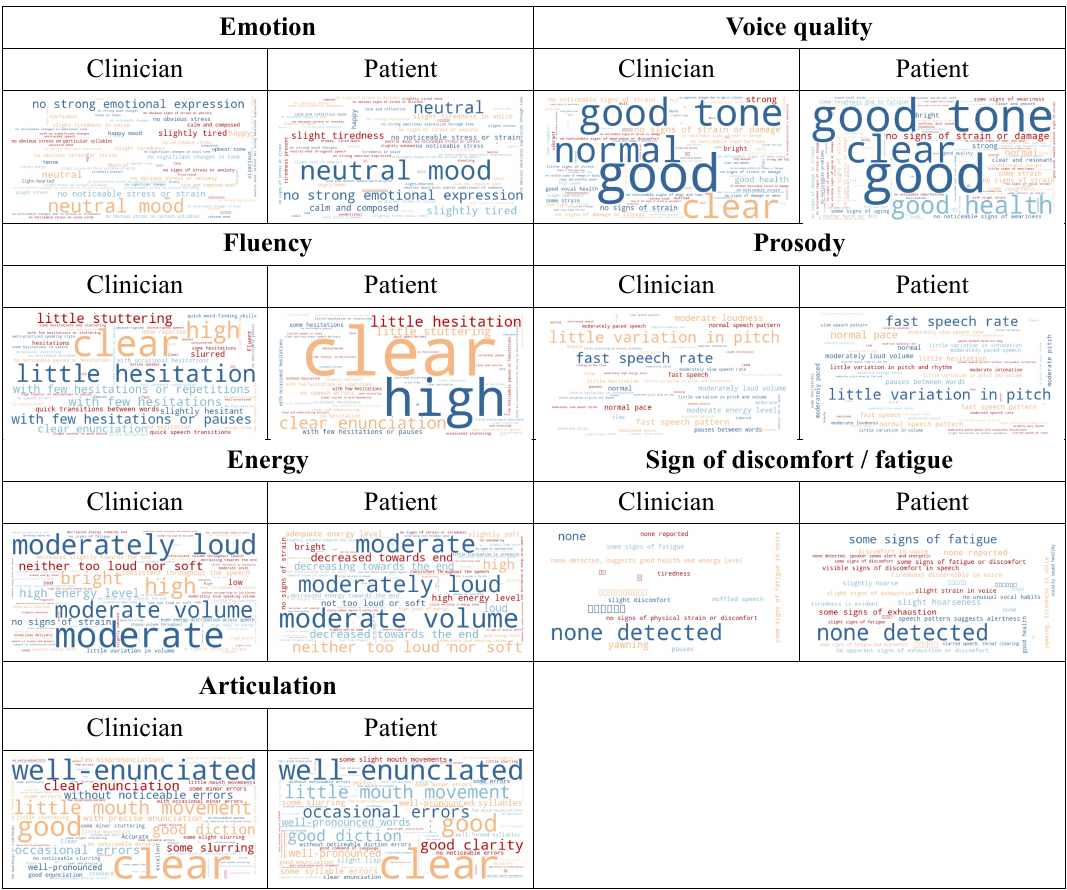} \hfill
  \caption {Wordcloud of Qwen2-Audio-7B-Instruct-generated phrase frequencies for clinician and patient voices.}
    \label{fig:qwen2-audio-wordcloud}
\end{figure}

\subsection{Analysis of Clinician and Patient Voices with Qwen2-Audio-7B-Instruct (Figure~\ref{fig:qwen2-audio-wordcloud})}

\newpage 
\subsection{Analysis of Patient Voices Across Health Conditions with GPT-4o-mini-audio (Table~\ref{tab:4omini-description})}

\begin{table}[htbp!]
    \centering
  \caption {Top-3 GPT-4o-mini-audio-generated  acoustic descriptions for patients per health condition group using TF-IDF analysis. The phrases within each group are ranked from most frequent to least frequent.
  }
  \includegraphics[width=\linewidth]{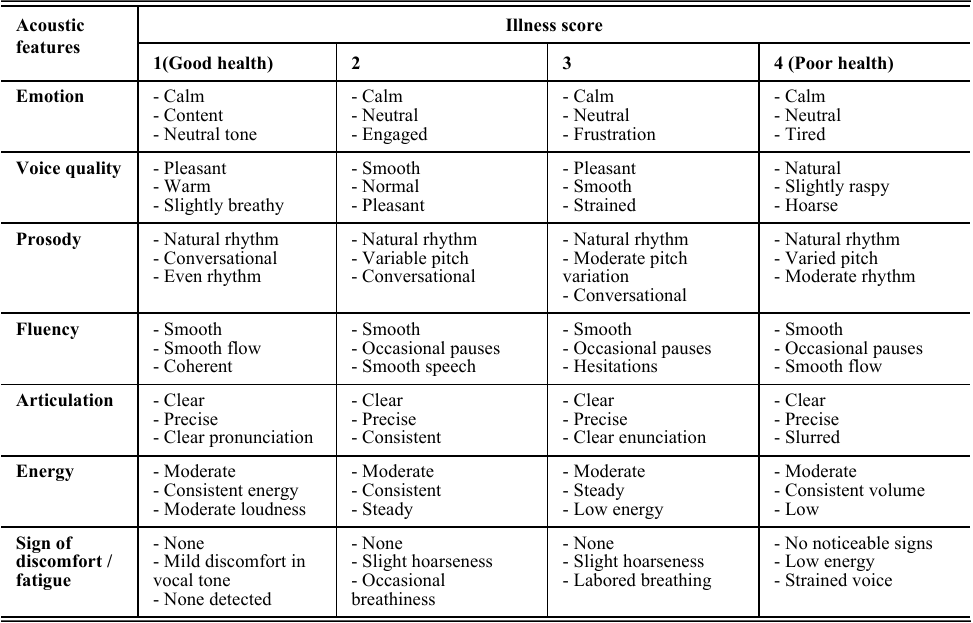} \hfill
    \label{tab:4omini-description}
\end{table}

\newpage

\subsection{Analysis of Patient Voices Across Health Conditions with Qwen2-Audio-7B-Instruct (Table~\ref{tab:qwen2-7b-description})}

\begin{table}[htbp!]
    \centering
  \caption {Top-3 Qwen2-Audio-7B-Instruct-generated acoustic descriptions for patients per health condition groups using TF-IDF analysis. The phrases within each group are ranked from most frequent to least frequent.
  }
  \includegraphics[width=\linewidth]{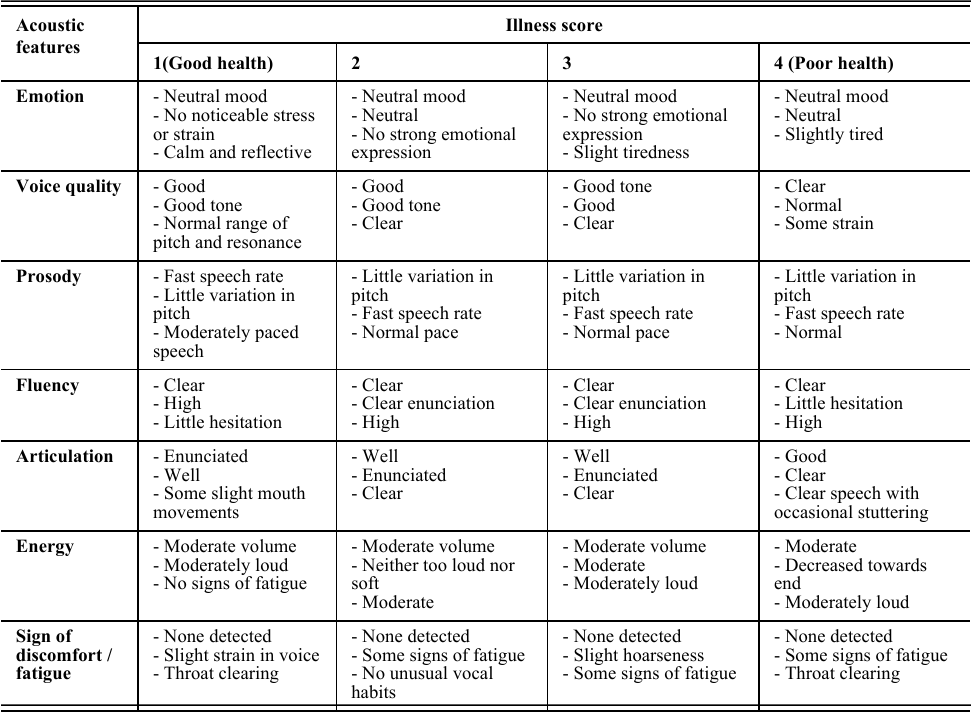} \hfill
    \label{tab:qwen2-7b-description}
\end{table}

\end{document}